# Erasing and Correction of Liquid Metal Printed Electronics Made of Gallium Alloy Ink from the Substrate

Rongchao Ma [1], Yixin Zhou [1], and Jing Liu[1,2*]

1. Key Lab of Cryogenics and Beijing Key Lab of CryoBiomedical Engineering,

Technical Institute of Physics and Chemistry,

Chinese Academy of Sciences, Beijing 100190, P. R. China

2. Department of Biomedical Engineering,

School of Medicine, Tsinghua University, Beijing 100084, P. R. China

**\*Address for correspondence:**

Dr. Jing Liu

Key Lab of Cryogenics,

Technical Institute of Physics and Chemistry,

Chinese Academy of Sciences,

Beijing 100190, P. R. China

E-mail address: jliu@mail.ipc.ac.cn

Tel. +86-10-82543765

Fax: +86-10-82543767

**Abstract**

Gallium-based liquid metals have recently been found important in a variety of newly emerging applications such as room temperature metal 3D printing, direct writing of electronics and biomedicine etc. In all these practices, one frequently encounters the situations that a printed circuit or track needs to be corrected or the unwanted parts of the device should be removed as desired. However, few appropriate strategies are currently available to tackle such important issues at this stage. Here we have identified several low cost ways toward this goal by comparatively investigating three typical strategies spanning from mechanical, chemical, to electrochemical principles, for removing the gallium-based liquid metal circuits or thin films. Regarding the mechanical approach, we constructed an eraser for removing the liquid metal thin films. It was shown that ethanol ($CH_3CH_2OH$) could serve as a good candidacy material for the mechanical eraser. In the chemical category, we adopted alkalis and acids to remove the finely printed liquid metal circuits and sodium hydroxide (NaOH) solution was particularly revealed to be rather efficient in making a chemical eraser. In the electrochemical strategy, we applied a 15 V voltage to a liquid metal thin film (covered with water) and successfully removed the target metal part. These methods were comparatively evaluated with each of the merits and shortcomings preliminarily clarified in the end. The present work is expected to be important for the increasing applications of the liquid metal enabled additive manufactures.

# 1 Introduction

Gallium or its alloy has low melting points, small vapor pressures (<$10^{-6}$ Pa at 500 ℃), and are generally non-toxic [1]. These merits enable them to be rather promising candidates for replacing mercury-based applications. For example, in biomedical field, the gallium-based liquid metals have been found important in a number of emerging areas, such as dental filling [2], injectable medical electronics [3], angiography [4], and bone repairing [5]. On the other hand, the gallium-based liquid metals are being increasingly tried in direct writing electronics or printed sensors [6-10], stretchable electronics [11], chip coolant [12], mini pumps [13], thermometer, electrodes [14], and so on. Recently, it is also found that the liquid metals possess unique ability of controllable transformation and movement under external electric fields [15].

However, any applications related to the liquid metals would encounter the problems of reparation, cleaning, or replacement. For example, some part of a liquid metal in a printed electronic circuit may require modification or just removal. Further, the liquid metal residue in the experiments and applications should be cleaned or recycled, and the metal parts in biomedical applications may need replacement. In all these circumstances, one has to find out a suitable way and material (or eraser) to remove or collect the unwanted liquid metals whenever needed. This requires one to either reduce the wettability of the liquid metal, or destroy the liquid metal directly [16]. However, a detailed literature on the erasing methods for the liquid metals is unavailable up to now. From the utilization aspect, it is highly desirable to establish appropriate ways for the coming tremendous needs.

In this study, we are dedicated to present the feasible methods for erasing the gallium-based liquid metal thin films or circuits. The methods are fundamentally rooted in three basic categories, i.e., mechanical, chemical, or electrochemical principles. The merits and disadvantages of each approach were comparatively evaluated.

# 2 Principle and Method

The experiments were carried out using liquid EGaIn (eutectic gallium indium) alloy, which has contents of 75.5% Ga and 24.5% In. The melting point of this alloy is 15.5 ℃ [8]. The gallium and indium used in the experiments have purities of 99.99%. The samples were prepared in the form of thin films and circuits.

The EGaIn thin films were written on glass substrates (75 mm×52 mm×1 mm) by the so-called "direct writing" method [7]: heat the substrates up to 45 - 50 ℃, and then paint the EGaIn onto the substrates with a brush or glass rod. The EGaIn was repeatedly painted and written to increase its oxide content, which can enhance the EGaIn's wettability. With increasing wettability between the EGaIn and substrates, the EGaIn finally well adheres to the substrates.

The EGaIn circuits were written (and printed) on PVC substrates via a liquid metal ballpoint pen [8,9]. The PVC substrates were used because they have better wettability than other substrates with the gallium-based liquid metals. This ensures that the EGaIn circuits can be successfully printed on the substrates. Furthermore, the PVC substrates can resist alkalis and acids.

It is known that the successful printing of a liquid metal circuit or the fabrication of a liquid metal thin film relies on the good wettability between the liquid metal and substrate. To remove the liquid metal circuits or thin films from the substrate, therefore, one has to either reduce the liquid metal's wettability or destroy the liquid metal directly. For this purpose, we proposed and tested three different representative approaches: mechanical, chemical, and electrochemical methods, respectively.

In the mechanical method, we constructed a liquid metal easer to remove the EGaIn thin films written on glass substrates. In the chemical method, we removed the EGaIn circuits using sodium hydroxide (NaOH) solution and hydrochloric acid (HCl) solution, respectively. The concentrations of these solutions are 6 mol/L. Regarding the electrochemical method, we removed the EGaIn thin films using a current/voltage power supply, which has a rated voltage of 20 V.

To observe the details of the erasing process, we used an optical microscope to record the images. Because the melting point of the EGaIn alloy is 15.5 ℃, we carried out all the measurements at 24-26 ℃ to ensure that the EGaIn alloy is in liquid state.

## 3 Results and Discussion

We have comparatively investigated three typical approaches for removing the liquid EGaIn thin films and circuits: mechanical, chemical, and electrochemical methods. The experimental results are as follows.

### 3.1 Mechanical method

The mechanical method (or physical method) here means a way to remove the liquid metals using external mechanical force, which does not rely on any chemical reactions. This method requires various mechanical scrubbing processes. Therefore, a critical step in such method is to prevent the liquid metal from sticking to the substrate again during the scrubbing process. From this point, we have built up a liquid metal eraser.

Fig. 1A shows the geometrical structure of the eraser, which is consisted of a remover container and cotton head. The cotton head is used for scratching the liquid metal. The remover in the container can be successfully conveyed to the cotton head along the cotton fibers. Fig. 1B shows an as-grown liquid metal thin film used in the experiment, which was eutectic gallium indium (or EGaIn) on a glass substrate (75mm×52 mm) prepared by the "direct writing" method [7]. One can see that, when the EGaIn thin film was scratched by the eraser without soaking ethanol, the EGaIn thin film cannot be completely cleaned (see Fig. 1C). However, when the EGaIn thin film was scratched by the eraser filled with ethanol, the EGaIn thin film can be easily removed (see Fig. 1D).

The above results show that the remover (ethanol) plays an important role. Therefore, it is critical to choose a suitable remover in the mechanical method. To this end, we have tested a number of typical room temperature liquid materials (see Table 1). By comparing the various liquid materials, we found that ethanol is a good candidate because it is nontoxic, fast drying, and can remove the gallium-based liquid metals efficiently.

**Table 1** Removers for mechanical method

| Name | Molecular formula | Melting point (°C) | Viscosity (cP @25 °C) | Toxic |
|---|---|---|---|---|
| Water | $H_2O$ | 0 | 0.894 | No |
| Ethanol [1] | $CH_3CH_2OH$ | -114 | 1.074 | No |
| Methanol [1] | $CH_3OH$ | -97.6 | 0.544 | Yes |
| Acetone [1] | $CH_3CHO$ | -95 | 0.306 | Yes |
| Formic acid [1] | HCOOH | 8.40 | 1.57 | Yes |
| Benzene [1] | $C_6H_6$ | 5.53 | 0.604 | Yes |

The reason that a remover can help in removing liquid metals could be explained as follows: when an eraser runs on the surface of a liquid metal thin film, the fibers in the cotton head create a number of scratches in the liquid metal thin film. But these scratches usually can self-cure due to the strong wettability between the liquid metal and substrate. Therefore, the liquid metal cannot be efficiently removed when the cotton head is dry (without any remover). However, if the eraser is filled with remover, the remover can fill the scratches faster than liquid metal does because the remover has a higher fluidity than that of the liquid metal. Therefore, the scratches cannot be cured anymore and the liquid metal thin film is cut into small pieces. The small pieces are further cut by the fibers and pushed forward to form more-or-less spheroidal particles. Finally, the liquid metal particles are covered by the remover and cannot wet the substrate anymore. The liquid metal particles are also isolated by the remover and cannot combine into larger droplets. In this sense, the remover can prevent the liquid metal from sticking to the substrate and help in removing the liquid metal.

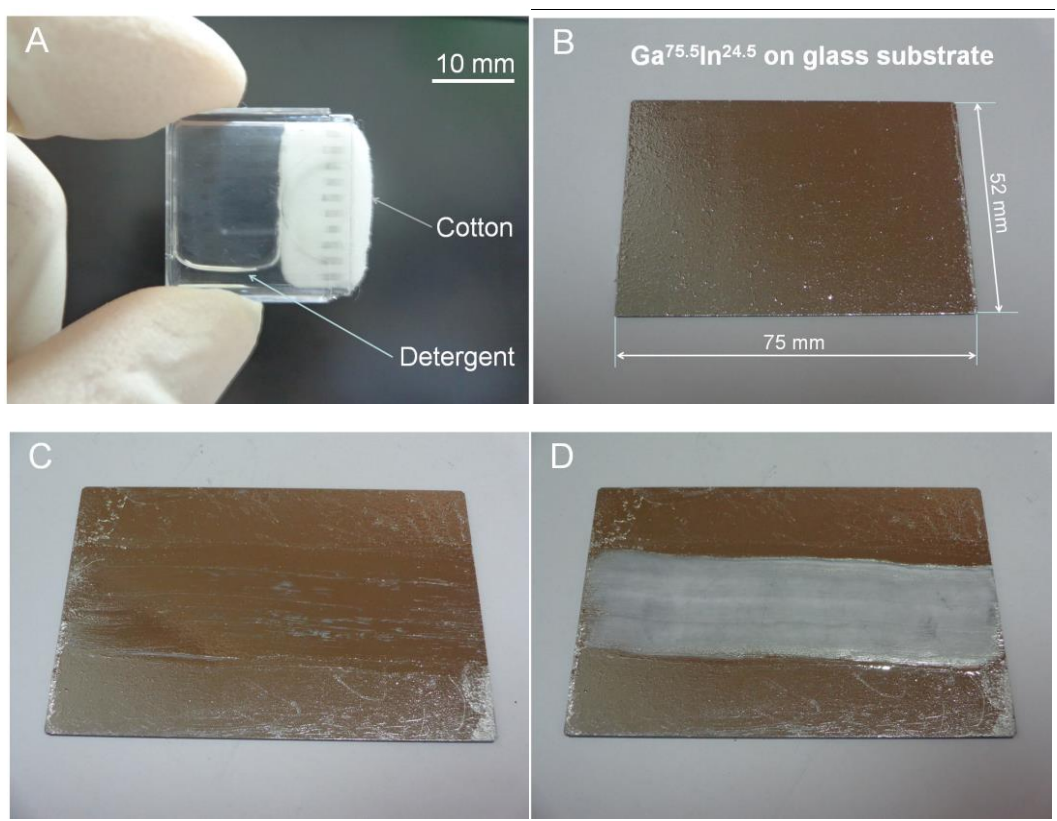

**FIG. 1.** Mechanical method for removing EGaIn (eutectic gallium indium, 75.5% Ga and 24.5%) thin films. (a) A liquid metal eraser consisted with a remover container and cotton head. (b) An EGaIn thin film prepared on a glass substrate (75mm×52mm×1 mm). (c) The EGaIn thin film was scratched by the eraser without ethanol. The EGaIn thin film cannot be efficiently removed. (d) The EGaIn thin film was scratched by the eraser with ethanol. The middle part of the EGaIn thin film was successfully removed.

The above analysis indicates that the erasing effect could be determined by: (1) the remover's viscosity, (2) the wettability between the remover and liquid metal. This means that the mechanical method strongly relies on choosing a suitable remover. A good remover should have a low viscosity and good wettability with the liquid metal. This is because:

(1) If the remover's viscosity is higher than that of the gallium-based liquid metal, then the following problems may come up: (a) the remover cannot fill the scratches promptly because the remover has worse fluidity than that of liquid metal; (b) the remover cannot wrap up the liquid metal particles because the liquid metal has better fluidity and can leak out quickly; (c) the eraser cannot smoothly slide on the surface because the remover may stick to the substrate and it is also hard to remove the remover.

(2) A good remover must be wettable to the liquid metal and finally can encase the liquid metal particles, but the remover's wettability with the substrate is not critical. One may have a better understanding on this by comparing to soap: to remove the dirt on cloth, the soap molecules have to wrap the dirt molecules, but it does not matter whether the soap is wettable to the cloth. One can further see this from a pencil eraser [17]: the interaction between the eraser and the graphite of the pencil is stronger than the interaction between the paper and graphite. Therefore, the pencil eraser can absorb the graphite and remove it (the pencil marks) after a number of scrubbing processes. But the pencil eraser should not "wet" the paper. Otherwise, the pencil eraser scraps will stick on the paper.

### 3.2 Chemical method

In this section, we have tested the possibility of using a chemical method to remove the liquid metal circuit. The basic idea is to either choose a chemical material to react with the gallium oxides and then reduce the wettability between liquid metal and substrate, or directly react with (destroy) the liquid metal. The chemical materials adopted should be nontoxic, environment friendly, do not react with (etch) the substrate, and do not affect the neighboring circuit. Finally, the remainder must be easy to remove.

It is known that gallium and its alloys usually do not wet most of materials. But they become wettable to most of the materials when they are partially oxidized [10, 18-23]. This means that the gallium's wettability is strongly related to its oxidizations. For this reason, the gallium-based liquid metals used in printable electronics are usually partially oxidized in advance to increase its wettability

[6-10].

It is also known that, similar to aluminum [1], gallium is a two-fold material which reacts with both alkalis and acids. Therefore, one can remove the gallium-based alloys using alkali solutions or acid solutions.

**(i) Alkalis**

The commonly used soluble alkalis include sodium hydroxide (NaOH) and potassium hydroxide (KOH). Here we choose one of the best erasing detergents, the NaOH solution as an example to demonstrate how to apply an alkali to remove the gallium-based liquid metals.

A NaOH solution with a concentration of 6 mol/L was prepared in the experiments. The chemical equations for the reaction of NaOH with gallium oxides, and NaOH with gallium are as follows:

$$Ga_2O_3 + 2\ NaOH \rightarrow 2\ NaGaO_2 + H_2O \tag{1}$$

$$2\ Ga + 2\ NaOH + 2\ H_2O \rightarrow 3\ H_2 + 2\ NaGaO_2 \tag{2}$$

$$2\ Ga + 6\ NaOH + x\ H_2O \rightarrow 3\ H_2 + 2\ Na_3GaO_3 + x\ H_2O \tag{3}$$

$$2\ Ga + 2\ NaOH + 6\ H_2O \rightarrow 3\ H_2 + 2\ NaGa(OH)_4 \tag{4}$$

Fig. 2A shows the photographs of the liquid metal circuits taken under an optical microscope. The circuits were prepared with EGaIn on a PVC substrate via a liquid metal ballpoint pen [9]. Fig. 2B indicates that a drop of NaOH solution was dropped onto the middle circuit from a syringe needle. The EGaIn circuit was broken immediately and then contracts quickly with increasing time (see Fig. 2C and Fig. 2E). After 20 seconds, the middle circuit contracts to the ends (see Fig. 2E). After drying the NaOH solution, one can clearly see that the middle EGaIn circuit has been completely removed (see Fig. 2F).

The results show that the 6 mol/L NaOH solution can easily remove the printed liquid EGaIn circuit. The reason lies in that the alkalis can react with gallium oxide and then reduce the wettability between the circuit and substrate. Thereafter, the liquid metal EGaIn circuit contracts and is then removed from the PVC substrate.

**(ii) Acids**

From chemistry, we know that gallium and its oxidations also react with strong acids, including hydrochloric acid (HCl), perchloric acid ($HClO_4$), hydrobromic acid (HBr), hydriodic acid (HI), nitric acid ($HNO_3$), and sulfuric acid ($H_2SO_4$). This means that one can also use an acid to remove the gallium-based liquid metals. Here we used HCl as an example to demonstrate how to use an acid solution to remove the gallium-based liquid metals.

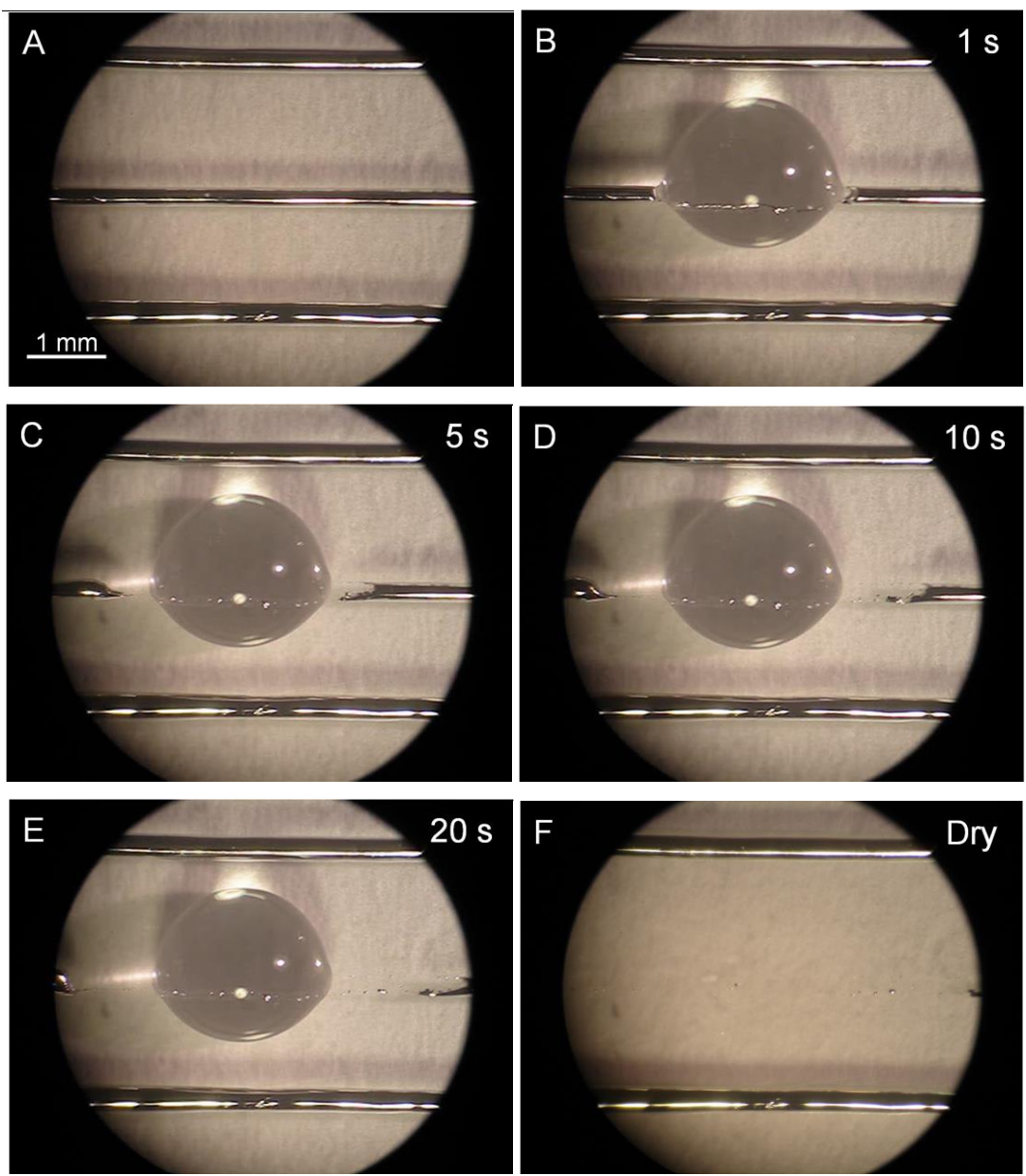

**FIG. 2.** Chemical method for removing liquid EGaIn circuits using sodium hydroxide (NaOH) solutions. (a) The as-grown liquid EGaIn circuits prepared on a PVC substrate. (b) A drop of 6 mol/L NaOH solution was dropped onto the middle circuit from a syringe needle. (c)-(e) The photographs of the EGaIn circuits after the NaOH solution was dropped onto the middle circuit for 5, 10, and 20 seconds, respectively. (f) The NaOH solution was dried with cotton and the middle circuit was completely removed.

A HCl solution with a concentration of 6 mol/L was prepared in the experiments. The chemical equations between the HCl and gallium oxides, gallium are as follows:

$$Ga_2O_3 + 6\ HCl \rightarrow 2\ GaCl_3 + 3\ H_2O \qquad (5)$$

$$2\ Ga + 6\ HCl \rightarrow 2\ GaCl_3 + 3\ H_2 \qquad (6)$$

Fig. 3A shows the photographs of the liquid metal circuits taken under an optical microscope. The circuits were prepared with EGaIn on a PVC substrate used a liquid metal ballpoint pen (similar to those used in alkalis experiments). Fig. 3B shows that a drop of HCl solution was dropped onto the middle circuit from a syringe needle. The EGaIn circuit was broken and then started to contract with increasing time (see Fig. 3C). After the HCl solution were dropped onto the EGaIn circuit for 10 seconds, the EGaIn circuit was completely removed (see Fig. 3D).

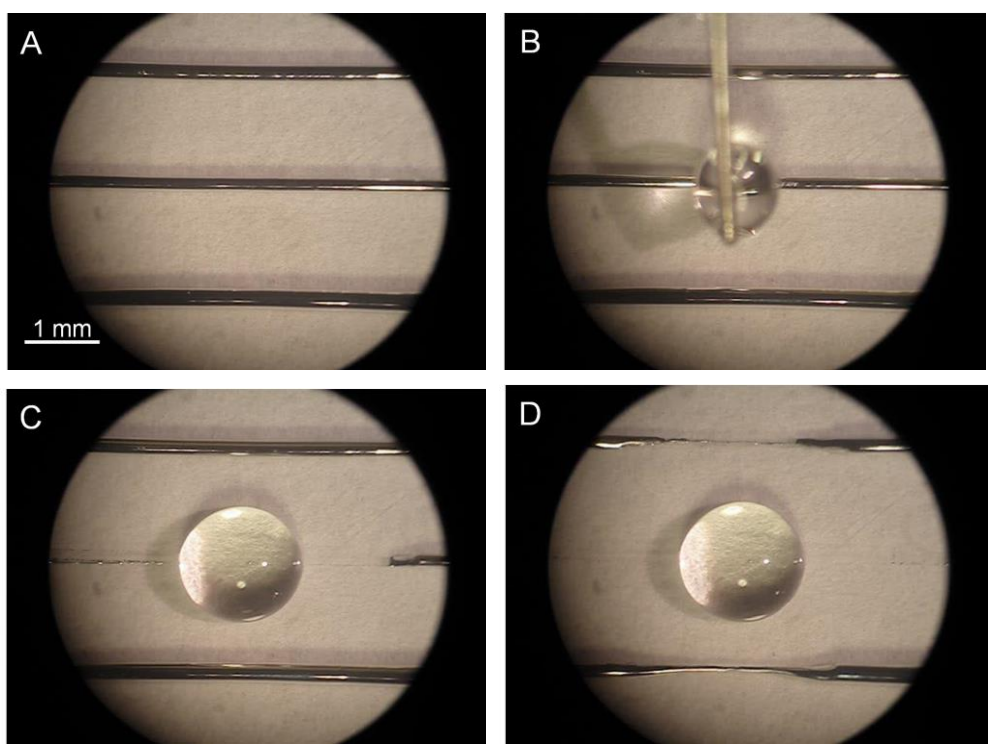

FIG. 3. Chemical method with a hydrochloric acid (HCl) solution for removing liquid EGaIn circuits. (a) The as-grown liquid EGaIn circuits prepared on a PVC substrate. (b) A drop of 6mol/L hydrochloric acid (HCl) was dropped on to the middle circuit from a syringe needle. (c) The liquid EGaIn circuit after the HCl solution was dropped onto the middle circuit for one second. (d) The liquid EGaIn circuit after the HCl solution was dropped onto the circuit for ten seconds. The neighboring circuits were corroded.

Fig. 3D further shows that the hydrochloric acid (HCl) not only etches the target circuit, but also causes the neighboring circuits to corrode. The same phenomenon was also observed in nitric acid ($HNO_3$). The reason for the occurrence of this phenomenon could be attributed to the volatility of the acids. In other words, the acids volatiles and drifts to the neighboring circuits, which also causes the neighboring circuits to corrode. This indicates that a volatile acid is not suitable for the use in removing the finely printed liquid metal circuits.

### 3.3 Electrochemical method

Based on above comparative tests and evaluations, we then turn to another way, the electrochemical method, to remove the liquid EGaIn metal thin films and circuits. This method is stimulated from a recently discovered phenomenon [15]: gallium-based liquid metal thin films contract automatically when covered by a solution with a voltage applied to the solution.

A current/voltage power supply with a rated voltage of 20 V was applied in the experiments. The electrochemical process may include the following chemical equation:

$$2\ Ga_2O_3 \rightarrow 4\ Ga + 3\ O_2 \qquad (7)$$

or

$$4\ Ga^{3+} + 12\ e^- \rightarrow 4\ Ga \qquad (8)$$

This equation shows that the electrochemical process can reduce gallium oxide to gallium. It indicates that the electrochemical process can reduce the wettability between the EGaIn thin film and substrate. Therefore, one can remove (or collect) the liquid metal thin films through the above electrochemical process.

Fig. 4A shows the photographs of a liquid EGaIn thin film taken under an optical microscope. The thin film was prepared with liquid EGaIn directly written on a glass substrate. In the measurements, the thin film was covered with a layer of water for the electrochemical process to occur. A 15 V voltage was then applied to the middle of the EGaIn thin film (Fig. 4B). The thin film between the two electrodes started to contract toward the cathode (see also Fig. 4C). Finally, the EGaIn thin film was removed after 6 seconds (see Fig. 4D).

The mechanism of the electrochemical method may be simply explained as follows: the electrochemical reaction reduces the gallium oxides between the liquid metal thin films and substrates. Consequently, it reduces the wettability between the gallium-based liquid metals and substrates. The liquid metal thin films then contract due to their strong surface tension.

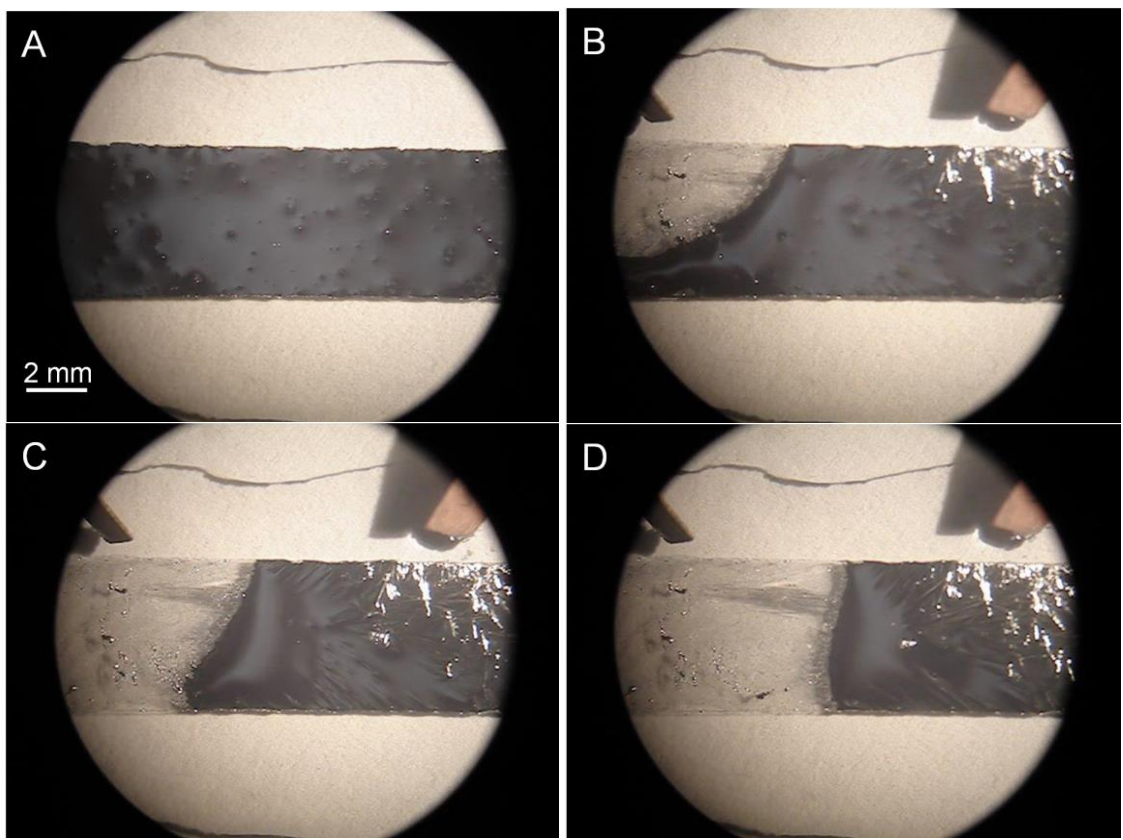

**FIG. 4.** Electrochemical method for removing a liquid EGaIn thin film. (a) A weakly adhered liquid EGaIn thin film was prepared on a glass substrate. The thin film was covered with a layer of water for the electrochemical process to occur. (b) The thin film start to contract toward the cathode after a 15 V voltage was applied to the thin film (after two seconds). (c) After the 15 V voltage was applied to the thin film for 4 seconds. (d) After the 15 V voltage was applied to the thin film for 6 seconds, and the thin film was removed.

## 4 Discussion

We have studied three representative types of methods for erasing gallium-based liquid metal thin films and circuits. Let us now compare the advantages and disadvantages of each method.

### 4.1 Technical advantage and disadvantage

The mechanical method relies on choosing a suitable remover. This indicates that one has more freedom in using this method to erase liquid metal. On the other hand, it is difficult to make an eraser with small geometrical size. Therefore, the mechanical method may be only suitable for removing the liquid metal thin films and circuits with large surface area, but not for removing the finely printed circuits. Furthermore, the mechanical strategy cannot remove solid thin films and solid circuits because this method requires that the liquid metal has good fluidity. At lower temperatures, however, this condition cannot always be satisfied because generally the gallium-based liquid metals have melting points above 7.6 ℃.

The chemical method can efficiently remove both liquid and solid thin films (and circuits). But the chemical method relies on chemical materials to remove the circuits. Considering that the chemical materials may also be corrosive to the substrates, one has to choose suitable chemical materials for certain substrates. For example, glass and silicon substrates react with alkalis, but not acids. Therefore, one needs to use acids, but not alkalis to remove the gallium-based liquid metal thin films written or printed on glass and silicon substrates. Finally, it should be also mentioned that the volatile acids, such as $HCl$ and $HNO_3$, cannot be used to remove the finely printed circuits because they cause the neighboring circuits to corrode due to its volatility.

The electrochemical method allows people to control the direction of the liquid metal contraction (contracts toward the cathode, see Fig. 4C). However, the erasing force generated in the electrochemical process is usually weak. Thus, the electrochemical method can only be used for removing weakly adhered liquid metal films and circuits, but not for firmly adhered thin films and solid circuits. Fortunately, this method can be combined with the former two erasing approaches to obtain better cleaning quality.

**4.2 Cost of material and equipment**

The mechanical method is simple and easy. It needs remover, but does not need any complicated devices. Therefore, the mechanical method is an economical method for erasing the liquid metal.

The chemical method needs consumables, like alkalis and acid. It also needs fine and corrosive-resistant equipment for transmitting the etchant. The cost is then higher than that of the mechanical method.

The electrochemical method does not need any consumables, but it needs a power supply for enabling the electrochemical reaction. The cost is also higher than that of the mechanical method, which may limit the application of the electrochemical method.

**4.3 Environmental/safety concerns**

The mechanical method relies on choosing a suitable remover. It does not bring in any environmental problems when using a nontoxic remover, such as water and ethanol. However, it may result in environmental/safety problems if one chooses the toxic removers (see Table 1).

The chemical method needs etchants which are strong alkalis and acid. These materials are corrosive and may result in environmental/safety problems.

The electrochemical method does not bring in any chemical contaminators (see chemical equation Eq.(7) amd Eq.(8)). Thus, it is environment friendly.

# 5 Conclusion

In summary, this study has investigated the feasible methods for erasing gallium-based liquid metal thin films and circuits. Three typical approaches were proposed and comparatively evaluated, i.e., mechanical, chemical, and electrochemical methods. Overall, the mechanical way is suitable for removing large liquid metal thin films and circuits, but has difficulty in removing finely printed circuits and solid circuits. Ethanol is a good remover for this purpose. The chemical method is rather efficient for cleaning finely printed circuits which are in either liquid or solid states. But it would bring in chemical contaminations. According to the experiments, sodium hydroxide (NaOH) solution could serve as a good etchant for this method. The electrochemical method is clean and environment friendly, but it can only remove the weakly adhered thin films or circuits. Future engineering approaches for better erasing the electronic circuits or objects made from the liquid metal additive manufacture can be enabled from these basic strategies or their combinations.

**Acknowledgments**

This work is partially supported by the China Postdoctoral Science Foundation (Grant No. 2013M541048) and Key Project Funding of Chinese Academy of Sciences and Beijing Municipal Science and Technology Funding under Grant No.Z151100003715002.

# References

[1] Haynes, W. M., 2014, CRC Handbook of Chemistry and Physics, CRC Press, 95th edition.

[2] Horasawa, N., Takahashi, S., Marek, M., 1999, "Galvanic interaction between titanium and gallium alloy or dental amalgam," Dental Materials, **15**, pp. 318–322.

[3] Jin, C., Zhang, J., Li, X. K., Yang, X. Y., Li, J. J., Liu, J., 2013, "Injectable 3-D fabrication of medical electronics at the target biological tissues," Scientific Reports, **3**, 3442-1-7.

[4] Wang, Q., Yu, Y., Pan, K., Liu, J., 2014, "Liquid metal angiography for mega contrast x-ray visualization of vascular network in reconstructing in-vitro organ anatomy," IEEE Trans. on Biomedical Engineering, **61**, pp. 2161-2166.

[5] Yi, L. T., Jin, C., Wang, L., Liu, J., 2014, "Liquid-solid phase transition alloy as reversible and rapid molding bone cement," Biomaterials, **35**, pp. 9789-9801.

[6] Boley, J. W., White, E. L., Chiu, G. T.-C., Kramer, R. K., 2014, "Direct writing of gallium-indium alloy for stretchable electronics," Adv. Funct. Mater, **24**, pp. 3501-3507.

[7] Gao, Y., Li, H., Liu, J., 2012, "Direct writing of flexible electronics through room temperature liquid metal ink," PLoS ONE, **7**, pp. e45485.

[8] Zheng, Y., He, Z., Yang, J., Liu, J., 2014, "Personal electronics printing via tapping mode composite liquid metal ink delivery and adhesion mechanism," Sci. Rep., **4**, pp. 4588.

[9] Zheng, Y., Zhang, Q., Liu, J., 2013, "Pervasive liquid metal based direct writing electronics with roller-ball pen," AIP Advances, **3**, pp.112117.

[10] Li, H., Yang, Y., Liu, J., 2012, "Printable tiny thermocouple by liquid metal gallium and its matching metal," Appl. Phys. Lett., **101**, pp. 073511.

[11] Park, J., Wang, S., Li, M., Ahn, C., Hyun, J. K., Kim, D. S., Kim, D. K., Rogers, J. A., Huang, Y., Jeon, S., 2012, "Three-dimensional nanonetworks for giant stretchability in dielectrics and conductors," Nature Communications, **3**, pp. 916.

[12] Deng, Y., Liu, J., 2010, "Design of practical liquid metal cooling device for heat dissipation of high performance CPUs," ASME J. Electron. Packag., **132**, pp. 031009.

[13]Gao, M., Gui, L., 2014, "A handy liquid metal based electroosmotic flow pump," Lab Chip, **14**, pp. 1866-1872.

[14] Surmann, P., Zeyat, H., 2005, "Voltammetric analysis using a self-renewable non-mercury electrode," Anal Bioanal Chem, **383**, pp. 1009-1013.

[15] Sheng, L, Zhang, J., Liu, J., 2014, "Diverse Transformations of Liquid Metals Between Different Morphologies," Advanced Materials, **26,** pp. 6036-6042.

[16] Wang, J., Liu, S., Guruswamy, S., Nahata, A., 2013, "Reconfigurable liquid metal based terahertz metamaterials via selective erasure and refilling to the unit cell level," Appl. Phys. Lett., **103**, pp. 221116.

[17] Petroski, H., 1992, The Pencil: A History of Design and Circumstance. New York: Alfred A. Knopf, 1st edition.

[18] Liu, T., Sen, P., Kim, C., 2012, "Characterization of nontoxic liquid-metal alloy galinstan for applications in microdevices," IEEE J. Microelectromechanical Systems, **21**, pp. 443-450.

[19] Gao, Y. X., Liu, J., 2012, "Gallium-based thermal interface material with high compliance and wettability," Appl Phys A, **107**, pp. 701–708.

[20] Regan, M. J., Tostmann, H., Pershan, P. S., Magnussen, O. M., DiMasi, E., Ocko, B. M., Deutsch, M., 1997, "X-ray study of the oxidation of liquid-gallium surfaces," Phys. Rev. B, **55**, pp. 10786-10790.

[21] Doudrick, K., Liu, S., Mutunga, Eva M., Klein, K. L., Damle, V., Varanasi, K. K., and Rykaczewski, K., 2014, "Different shades of oxide: From nanoscale wetting mechanisms to contact printing of gallium-based liquid metals", Langmuir, 30, pp. 6867−6877.

[22] Zhang, Q., Gao, Y. X., Liu, J., 2014, "Atomized spraying of liquid metal droplets on desired substrate surfaces as a generalized way for ubiquitous printed electronics", Applied Physics A, 116, pp. 1091–1097. (also see arXiv:1311.2158, 2013)

[23] MIT Technology Review, 2013, "Liquid Metal Printer Lays Electronic Circuits on Paper, Plastic, and Even Cotton", Nov. 19, 2013, access: https://www.technologyreview.com/s/521871/liquid-metal-printer-lays-electronic-circuits-on-paper-plastic-and-even-cotton/.